\documentclass[twocolumn]{article} 
\newcommand{\RRR}{$I\hspace{-0.25em}R^3$}
\newcommand{\ket}[1]{\vert#1\rangle} 

\begin{document}
\title{ UNVEILED REALITY: COMMENT ON D'ESPAGNAT'S\\
NOTE ON MEASUREMENT}
\author{Ulrich Mohrhoff\\ 
Sri Aurobindo International Centre of Education\\ 
Pondicherry-605002 India\\ 
\normalsize\tt ujm@satyam.net.in} 
\date{}
\maketitle 
\begin{abstract}
According to d'Espagnat we must choose between nonlinear breaks in quantum state 
evolution and weak objectivity. In this comment it is shown that this choice is forced on 
us by an inconsistent pseudo-realistic interpretation of quantum states. A strongly 
objective one-world interpretation of linear quantum mechanics is presented. It is 
argued that the weak objectivity favored by d'Espagnat is, in fact, inconsistent with 
quantum mechanics.
\end{abstract}

\section{\large CRITIQUE}

If both Alice and Bob see a teapot on the table then there is a teapot on the table. Or is 
there? In a recent article~\cite{dE01} Bernard d'Espagnat contrasts ``objectivistic realism'' 
(Option~A) with an alternative theory of science (Option~B) close to Putnam's ``internal 
realism''~\cite{PutnamRR}, which is close to Kant's theory of science. According to 
Option~A, the antecedent (Alice and Bob see a teapot) and the consequent (there is a 
teapot on the table) express two different states of affairs, such that the latter may be 
ontologically correlated, if not causally connected, with the former. According to 
Option~B, the consequent is equivalent to the conditional statement ``If Cecily would 
look to see what is on the table she would see a teapot.'' This leaves no room for an 
ontological correlation, let alone a causal link, between the antecedent and the 
consequent.

According to Option~A, the purpose of science is to discover how things are in 
themselves, rather than how they appear to us. According to Option~B, science aims to 
describe, as concisely as possible, the common (intersubjective or weakly objective) 
denominator of human experience. There are sound arguments in favor of either view. 
Science is driven by the desire to know how things {\it really} are. It owes its immense 
success in large measure to its powerful ``sustaining myth''~\cite{MerminSM}---the 
belief that this can be discovered. Neither the ultraviolet catastrophe nor the spectacular 
failure of Rutherford's model of the atom made physicists question their faith in what 
they can achieve. Instead, Planck and Bohr went on to discover the quantization of 
energy and angular momentum, respectively. If today we seem to have reason to 
question our ``sustaining myth'', it ought to be taken as a sign that we are once again 
making the wrong assumptions, and it ought to spur us on to ferret them out. A retreat 
from Option~A to Option~B should be seen for what it is---a cop-out.

Yet it takes only a millisecond's reflection to realize the naivity of the notion that the 
world as it is in itself, out of relation to human minds or brains, is just like the 
phenomenal world---the world as we humans perceive it. The first lesson of science it 
that appearances are deceptive. On Option~A, science peers beyond the phenomenal 
world at the world as it is in itself. But the notion that the world in itself is just like the 
world as we humans do (or eventually will) {\it conceive} it, seems just as naive. By 
definition, the world in itself has as little to do with human concepts and theories as it 
has with human perceptions. The ``primary qualities'' of Locke are as much products of 
the human mind as are the ``secondary'' ones.

In response to this line of reasoning one may point to ``the unreasonable effectiveness of 
mathematics in the natural sciences''~\cite{WignerUE} as the indicator of a kinship 
between the human mind and whatever is responsible for the structure of the world in 
itself. This again may be countered by pointing to our apparent inability to make sense 
of quantum mechanics (QM) as a sign that we are not all that well equipped mentally. 
However, these arguments are metaphysical, and of metaphysical arguments we ought to 
be wary, considering the ``unreasonable ineffectiveness of philosophy'' in the natural 
sciences pointed out by Weinberg~\cite{Weinberg}.

As a matter of fact, these metaphysical arguments are beside the point. The question is 
not whether the world in itself (assuming there is a way of making sense of this concept) 
is like the world as we conceive it. The question is not even whether QM compels 
physicists to forswear their ``sustaining myth''. As physicists we are committed to 
discovering such ways of thinking about what we experience as are consistent with 
Option~A. It is not within our purview to question our ``sustaining myth''. The question 
is not even whether unadulterated QM is consistent with Option~A. The question is, 
which way of thinking about unadulterated QM {\it is} consistent with Option~A? We 
need to discover whatever deep-seated {\it physical} (rather than metaphysical) 
misconceptions stand in the way of making sense of QM. It is {\it these} that we must 
disavow. To renounce Option~A in favor of Option~B is overkill.

The literature on measurement theory is so replete with mutually supportive 
inconsistencies that it is difficult for a critic to know where to begin. D'Espagnat rightly 
admonishes us to
\begin{quote}
be cautious when using the notion of `state', which\ldots has questionable metaphysical 
implications.
\end{quote}
He observes that
\begin{quote}
if\ldots we only worry about predicting what are our chances of observing this or that, 
and if, correlatively, we impart to the word `state' no other meaning than that of 
designating a mathematical tool allowing for such predictions, we meet with no 
ambiguities whatsoever.
\end{quote}
At the very least this should make one suspect that problems arising when quantum 
states are taken for more than algorithms for assigning probabilities to the possible 
results of possible measurements, are pseudo-problems.

That a quantum state is such a probability algorithm is evident from the minimal 
instrumentalist interpretation of QM, which constitutes the common denominator of all 
possible interpretations~\cite{Redhead}. It is equally evident from Jauch's {\it 
definition} of the ``state'' of a quantum-physical system as a probability measure 
resulting from a preparation of the system and his proof~\cite{Jauch}---based on 
Gleason's theorem~\cite{Gleason}---that every such probability measure has the 
well-known density-operator form, which reduces to the familiar Born probability measure if 
the density operator is idempotent. But if a quantum state is a probability algorithm, 
then it cannot also represent an actual state of affairs. How could it? A probability 
algorithm is one thing, an actual state of affairs belongs to an altogether different 
category. This immediately disposes of the ``measurement problem'' in its crudest form, 
which treats the state vector as an actual state of affairs with two modes of change, one 
governed by the Schr\"odinger equation and another governed by the projection 
postulate.

What changes in these two ways is a probability measure, and this for reasons that are 
obvious rather than mysterious. Probabilities are assigned, on the basis of relevant facts, 
to possible events or states of affairs (indicating possessed properties or values). They 
depend (i)~on the time $t$ of the events or states of affairs to which they are assigned 
and (ii)~on the facts on which they are based. They can therefore change not only in a 
``deterministic'' manner as functions of $t$ but also unpredictably with every new 
relevant fact. The successful completion of a measurement is the relevant fact {\it par 
excellence}. If the outcome of the measurement is unpredictable, as it generally is, it has 
to be included among the relevant facts on which probability assignments must 
subsequently be based. The outcome being unpredictable, the basis of relevant facts 
changes unpredictably as a matter of course, and so do the probabilities that we assign 
on this basis.

It seems to me that d'Espagnat does not sufficiently heed his own advice to ``be cautious 
when using the notion of `state'$\,$''. He refers to situations
\begin{quote}
in which the use of `realistic' sentences---involving the verbs `to have' and `to be'---is 
both harmless and convenient. This is when we know (for sure) beforehand that, if we 
measured an observable $B$ on a system $S$, we would get eigenvalue $b_k$ of $B$ as 
an outcome. In that case we may assert that system $S$ is in a state described by one of 
the eigenvectors of $B$ corresponding to eigenvalue $b_k$\ldots
\end{quote}
We may assert that the state $\ket{b_k}$ is associated with $S$, provided that all we 
mean by the ket $\ket{b_k}$ is the Born probability measures it defines. We may {\it not} assert 
that $\ket{b_k}$ {\it describes} the system and we may {\it not} assert that $S$ {\it is in a 
state}, for these phrases make no sense when applied to probability measures. Nor can 
we say that $B$ {\it has} the value $b_k$. This phraseology may be convenient but it is 
not by any means harmless. It totally obfuscates the ontological import of QM.

The first thing that needs to be understood about quantum-mechanical probabilities is 
that they are assigned to {\it conditional} statements. If QM assigns probability~1 to 
$b_k$ at a time $t$ then we are allowed to infer that a successful measurement of $B$ 
performed at the time $t$ will or would yield the result $b_k$. Thus, {\it if} $B$ is or 
were successfully measured at $t$ {\it then} $b_k$ will or would be found. The transition 
from this conditional statement to the blunt assertion that $B$ has the value $b_k$ at 
the time $t$ is illegitimate because the value of $t$ is defined by the measurement, as the 
time at which $B$ is measured, and is therefore {\it undefined} if the measurement is not 
actually made. Nothing warrants that blunt assertion except a matter of fact about the value of 
$B$ at the time $t$ (that is, an actual event or state of affairs that indicates the value of 
$B$ at $t$). As I have explained at length in Ref.~\cite{Mohrhoff00}, the (contingent) 
properties of quantum systems are {\it extrinsic} in the specific sense that they cannot be 
attributed unless they are indicated by facts. {\it No property is a possessed property 
unless it is an indicated property.} A position, in particular, does not exist for $S$ unless 
its possession by $S$ is indicated by an actual event or state of affairs. And the same 
holds true of the time at which a property is possessed. The time $t$ does not exist for 
$S$ unless it is the indicated time of possession (by $S$) of an indicated property~\cite{Time}.

The state vector $\ket{\psi(t)}$, and the probabilities it assigns to the possible 
results of possible measurements, depend on a time parameter~$t$. As long as probability 
assignments are based on the same set of relevant facts, its ``evolution'' is governed by 
the Schr\"odinger equation (or suchlike). What is the meaning here of ``evolution'', and what is the 
meaning of the parameter~$t$? Consider the Born probability $p\,(R,t)$ of finding a 
particle in a region $R$ at the time~$t$. While few would think of this probability as 
something that exists inside~$R$, many appear to think of it as something that exists at 
the time~$t$. The prevalent idea is that the possibility of finding the particle inside $R$ 
exists at all times, so the probability associated with this possibility also exists at all times 
and changes as a function of time. Yet the possibility that a property-indicating event or 
state of affairs happens or obtains at the time $t$ is not something that exists at the 
time~$t$, anymore than the possibility of finding the particle in $R$ is something that 
one can find inside~$R$. And the same, obviously, holds true of the probabilities 
associated with these possibilities.

Neither possibilities nor probabilities are things that subsist and change. To think of 
possibilities or probabilities as if they persisted and changed (``evolved'') in time 
(continuously and deterministically, as dictated by the Schr\"odinger equation) is a 
straightforward category error. It is this logical mistake that gives rise to the somewhat 
gentler avatar of the ``measurement problem'' which asks: How is it that during a 
measurement one of the persisting possibilities (or worse, one of the changing 
probabilities associated with them~\cite{Treiman}) becomes a fact, while the others 
cease to exist? A silly question, once you come to think of it, because possibilities aren't 
things that persist and probabilities aren't things that evolve. Saying in common 
language that a possibility becomes a fact is the same as saying that something that is 
possible---something that {\it can} be a fact---actually {\it is} a fact. How can that be a 
problem? This non-problem becomes a pseudo-problem if one forgets that there is only 
one kind of actuality and misconstrues the common-language ``existence'' of a possibility 
as a second kind of actual existence that can be converted into the genuine article by 
means of a measurement.

Since the probability $p\,(R,t)$ isn't something that exists at $t$ (anymore than it is 
something that exists in $R$), the parameter $t$ isn't the time at which the probability 
$p\,(R,t)$ exists. $p\,(R,t)$ isn't a thing of which we can say {\it when} it exists. {\it A 
fortiori} it isn't something that can evolve. Quantum-mechanical probability 
assignments are {\it conditional} on the existence of a matter of fact about the value of a 
{\it given} observable at a {\it given} time. $p\,(R,t)$ isn't associated with the possibility 
that all of a sudden, at the time~$t$, the particle ``materializes'' inside~$R$. It is the 
probability with which the particle is found in $R$, {\it given} that at the time $t$ it is 
found in one of a set of mutually disjoint regions (no matter which one, $R$ being one of 
them). The parameter $t$ on which this probability depends is the time of this actually 
or counterfactually performed position measurement. It refers to the time of a 
position-indicating event or state of affairs, without which it is utterly meaningless.

The above quotation---``if\ldots we only worry about predicting what are our chances of 
observing this or that''---creates the impression that the only choice we have in this 
regard is between (i)~thinking of quantum states exclusively as probability measures and 
(ii)~considering them also as warranting inferences to actual states of affairs (such as the 
possession of a property by a system), and that if we chose the former, we have nothing 
else to worry about. While option~(i) is, in fact, the only consistent way of thinking 
about quantum states, nothing could be further from the truth than the conclusion that 
there is nothing else to worry about. This conclusion is based on an erroneous 
identification of option~(i) with instrumentalism---the view that QM exclusively 
concerns statistical correlations between measurement outcomes, and that any attempt 
to go beyond these ``brute facts'' is idle metaphysics. There remains so much to worry 
about that we can't afford wasting time and effort over ``solving'' pseudo-problems arising 
from option~(ii).

Back to d'Espagnat's Note:
\begin{quote}
Similarly, when, as with the example of the stone on the path [or the teapot on the 
table], we know for sure that if we looked we would have the impression of seeing a 
certain physical system lying within a given region of space instead of outside it, we are 
allowed to consider this knowledge as enabling us to make some definite statements 
concerning the quantum mechanical description of this system. For instance, when the 
system in question is an electron we are allowed to infer from such a knowledge that the 
state vector of the electron (or, better to say, of the whole Universe including the 
electron) is an element of a certain set of vectors.
\end{quote}
Once again: There is no such thing as a quantum-mechanical {\it description} in the sense intended by 
d'Espagnat. All we know for sure is the statistical correlations that exist between property-indicating 
facts (diachronic correlations between the results of measurements performed on the 
same system at different times, and synchronic correlations between the results of 
measurements performed on different systems in spacelike separation). The task before 
us is to draw ontological inferences from these correlations while eschewing an 
inconsistent mathematical realism that interprets some or all of the mathematical 
symbols employed by QM as mirroring (representing, describing) the physical world, 
and that allows inferences from quantum states to possessed properties. If all we know 
for sure is that a position measurement, if one were successfully performed, {\it would} indicate 
the presence of an 
electron in a finite region $R$, we are {\it not} allowed to infer that the electron {\it is} 
in~$R$. Nor are we allowed to infer (from this knowledge) that a state vector (rather 
than an ``impure'' density operator) can be assigned to the electron. Nor is there a way 
of making sense of the state vector of the universe. The values of quantum-mechanical 
observables being extrinsic properties, they obviously cannot be attributed to the 
universe as a whole.

D'Espagnat fails to make due allowance for the difference between {\it mathematical} 
realism (in the sense just defined) and an ``objectivistic realism'' (to stick to his term) 
that (i)~allows perceptions of teapots to be correlated with the existence of real teapots 
and (ii)~holds that science is in the business of discovering the truth about things in 
themselves (including teapots). If mathematical realism were the only possible 
objectivistic realism then Option~B would quite arguably be the only way of making 
sense of QM. But it isn't.

The key issue is to find a way of thinking about quantum-mechanical correlations that is 
consistent with the existence of classical behavior at macroscopic scales, and a common 
way of dodging this issue is to implicate human consciousness or 
knowledge~\cite{MohrhoffFF}. This may have started with von Neumann's principle of 
psychophysical parallelism~\cite{vN}, according to which subjective perceptions 
correspond to objective (neural) processes. As far as I can tell the principle is sound, but 
it tends to be used in the wrong direction, by arguing from the definiteness of 
observation reports to some sort of superselection rule. It ought to be used instead to 
eliminate all references to observations (qua conscious and/or intentional acts). Nothing 
but confusion is created by dragging the mysterious relation between things and 
perceptions of things into discussions of QM. This relation has nothing to 
do with QM, for it exists between mental representations and facts and is anything but 
statistical, whereas QM concerns the statistical correlations that exist between facts and 
facts. (The relation between things and perceptions, by its very nature, doesn't even fall 
within the purview of the objectivistic paradigm, but precisely because it has nothing to 
do with QM, it doesn't follow that QM is inconsistent with Option~A.) 

D'Espagnat accepts the common point of view according to which
\begin{quote}
the Schr\"odinger time evolution leads, for the overall system $S$ composed of $S$ and 
the pointer\ldots to a state that is a superposition of macroscopically distinct states; a 
result which is incompatible with Option~A.
\end{quote}
The Schr\"odinger equation leads to nothing of this sort. It statistically correlates 
property-indicating facts (preparations not of systems but of probability measures) 
with property-indicating facts (measurement outcomes). It governs not the time 
evolution of a state (in the common-language acceptation of ``state'') but the dependence 
of probability measures on the time span between preparation and measurement. (Note 
that ``preparation'' and ``measurement'' can be exchanged: Diachronic correlations can 
be used to retrodict as well as predict, just as synchronic correlations can be used to 
assign probabilities to the possible results of Bob's measurement on the basis of the 
result of Alice's measurement as well as vice versa. Our sense of a directed time ``flow'' 
is irrelevant to the interpretation of QM~\cite{MohrhoffRK}.)

A ``superposition of states'' is a probability measure expressed in a form in which the 
probability amplitudes associated with the possible results of a certain measurement are 
explicit. It entails nothing but the wholly unproblematic common-language ``existence'' 
of several possibilities. An inconsistency between ``superpositions of states'' and the 
definiteness of observation reports only arises if one commits the category mistake of 
thinking of possibilities or probability measures as if they were facts or as if they 
entailed anything factual. Then one has to do some explaining. And the first question 
that arises is: Are wave function collapses in the mind but not in the world, or in the 
mind because they are in the world, or in the mind and therefore in the world? There 
aren't any wave function collapses, but if one must choose then d'Espagnat's conclusion 
appears inescapable: If they are in the mind because they are in the world then ``in a 
way or another the linear nature of the dynamics must be broken''~\cite{BG}. I heartily 
agree with d'Espagnat that
\begin{quote}
none of the schemes that materialize the break is as yet considered, for various reasons, 
as being fully convincing.
\end{quote}
But this isn't the fault of Option~A and it doesn't entail Option~B. None of those 
schemes is convincing because there isn't any ``break'' that needs to be ``materialized''. 
The actualization of a possibility is not a physical process~\cite{noteAP}.

If adulterations of QM are rejected, the choice is between (i)~only in the mind and 
(ii)~in the mind and therefore in the world. The first option leads to the many-worlds (or 
many-minds) extravaganza, the latter leads, credibly enough, to Option~B. If the 
premise is that system $S$ enters into a ``state of entanglement'' with apparatus $A$, 
then apparatus $A$ enters into a ``state of entanglement'' with Cecily's brain as she 
takes cognizance of the measurement outcome, and then the definiteness of observation 
reports combined with the principle of psycho-physical parallelism spells collapse. Since 
it would be a bit rich to claim that Cecily's peep at the pointer causes a collapse ``out 
there'' in the real world, divorcing the subject matter of science from the real world 
``out there'' seems the proper thing to do. If science is concerned only with the 
intersubjective (weakly objective) reality of the ``world as we know it'' (to use Kant's 
phrase), mind-induced collapses make more sense.

But this doesn't mean that they make sense. A ``state of entanglement'' is a probability 
measure just like any old quantum state, except that it gives {\it joint} probabilities for 
the possible results of possible measurements on more systems than one. If a probability 
measure changes unpredictably, it does so for reasons that are obvious rather than 
mysterious, as explained above. If anything gets entangled, it is possibilities, and if 
anything gets correlated, it is probabilities. There isn't any way of making sense of an 
{\it actual} ``state of entanglement''. Option~B therefore is just another gratuitous 
solution to a pseudo-problem. There is nothing wrong with the philosophy behind 
Option~B, except that it is irrelevant.
\begin{quote}
According to this trend of thought (considered as being the most reasonable one by, 
perhaps, the majority of contemporary philosophers), the fact that we perceive such 
``things'' as macroscopic objects lying at distinct places is due, partly at least, to the 
structure of our sensory and intellectual equipment.
\end{quote}
True enough but, as I said, beside the point, for the real issue is the ontological import 
of statistical correlations between facts and facts. The nonstatistical correlations between 
facts and perceptions have nothing to do with it. (I would also contest the parenthetical 
claim. Putnam, the erstwhile champion of internal realism~\cite{PutnamRR}, for one, 
now considers the same philosophy ``fatally flawed''~\cite{PutnamHC}.)

I grant d'Espagnat that dictionaries ``define'' facts in terms that reek of Option~B. {\it 
The Concise Oxford Dictionary} (8th edition, 1990) ``defines'' ``fact'' as a thing that is 
known to have occurred, to exist, or to be true; a datum of experience; an item of 
verified information; a piece of evidence. How else could it be ``defined''? ``Fact'', like 
``existence'', like ``reality'', is so fundamental a concept that it simply cannot be defined. 
So what is the editor of a dictionary to do? The obvious thing is to fall back on the 
metalanguage of epistemology. This shifts the burden of definition to such terms as 
``experience'' or ``knowledge''. Let's look them up: Experience is an ``actual observation 
of or practical acquaintance with {\it facts} or events''. Knowledge is ``awareness or 
familiarity gained by experience (of a person, {\it fact}, or thing)'' (italics supplied). 
Which shows, if anything, that such terms as ``experience'' or ``knowledge'' cannot be 
invoked to give meaning to the word ``fact''.

If ``fact'' is so fundamental a term that it cannot be defined, the existence of facts---the 
factuality of events or states of affairs---cannot be accounted for, any more than we can 
explain why there is anything at all, rather than nothing. (If something can be accounted 
for, it can be defined in terms of whatever accounts for it.) Before the mystery of 
existence---the existence of {\it facts}---we are left with nothing but sheer 
dumbfoundment. In spite of this, measurement theorists are busy trying to explain the 
emergence of facts (a.k.a. ``classicality''). The apparent need for this wholly gratuitous 
endeavor arises if one thinks of the possibilities to which QM refers, and/or of the 
probabilities it assigns to them, as if they constituted a self-existent matrix from which 
facts emerge---another category error due to taking the quantum state for more than a 
probability measure on the possible results of possible measurements.

Classical physics deals with nomologically possible worlds (that is, worlds consistent 
with physical theory). The question as to which of these worlds is real (agrees with the 
actual world) is of historical rather than scientific interest. Giving an answer to this 
question is strictly a matter of observation. Does this imply that classical physics makes 
sense only within a theory of science committed to Option~B? Obviously not. In 
classical physics the actual course of events is in principle fully determined by the 
actual initial conditions (or the actual initial and final conditions). In quantum physics 
it also depends on unpredictable actual events at later (or intermediate) times. Hence 
picking out the actual world from all nomologically possible worlds requires 
observation not only of the actual initial conditions (or the actual initial and final 
conditions) but also of those unpredictable actual events. Does this imply that quantum 
physics makes sense only within a theory of science committed to Option~B? If the 
answer is negative for classical physics, it is equally negative for quantum physics.

QM concerns statistical correlations between facts, and the correlations warrant 
interpreting the facts as indicative of properties. That is, they warrant the existence of 
a physical system or systems to which the indicated properties can be attributed. 
Suppose that we perform a series of position measurements, and that every position 
measurement yields exactly one result (that is, each time exactly one detector clicks). 
Then we are entitled to infer the existence of an entity that persists through time (if not 
for all time), to think of the clicks given off by the detectors as matters of fact about the 
successive positions of this entity, to think of the behavior of the detectors as position 
measurements, and to think of the detectors as detectors. If instead each time exactly 
two detectors click, we are entitled to infer the existence of two entities or, rather, of a 
physical system with the property of having two components. This property is as 
extrinsic as are the measured positions. There is a determinate number of entities only 
{\it because} every time the same number of detectors click. Not only the properties of 
things but also the number of existing things supervenes on the facts.

This ontological dependence of the properties and the number of things on facts warrants 
the distinction between two domains, a ``classical domain'' of facts and a ``quantum 
domain'' of properties that exist only because they are indicated (by facts). The 
impossibility of attributing to the properties of the quantum domain an intrinsic 
existence (or, equivalently, the necessity of attributing their existence to the classical 
domain) combined with the apparent impossibility of understanding the relation 
between the two domains within the objectivistic paradigm, is the reason behind the 
frequent invocation of consciousness in general and of Option~B in particular. If 
(i)~the real world is the quantum domain, and if (ii)~the properties of the quantum 
domain depend on the classical domain, and if (iii)~the classical domain can be neither 
defined nor accounted for by the quantum domain, then the conclusion that the 
classical domain is grounded in human experience is inescapable.

The second antecedent is certain. QM presupposes facts from beginning to end---from 
the preparation of a probability measure to a measurement. If the first antecedent is 
accepted, the third means that the classical domain isn't part of the real world, and this 
leads to the conclusion that the classical domain is ``in the mind''. On this view saying 
that the properties of the quantum domain exist only because they are indicated by 
what happens or is the case in the classical domain, is the same as saying that the 
properties of the world exist only because they are perceived---{\it esse est percipere aut 
percipi}. But the real world isn't the quantum domain. The real world is the classical 
domain plus whatever properties of the quantum domain can be inferred from the 
goings-on in the classical domain. The third antecedent remains true, inasmuch as the 
properties of the quantum domain are defined and accounted for by the properties of 
the classical domain rather than vice versa. But it doesn't mean that the classical 
domain isn't part of the real world. And therefore it doesn't follow that the classical 
domain is ``in the mind''.

\section{\large STRONGLY OBJECTIVE\\
ONE-WORLD INTERPRETATION\\
OF LINEAR QM}

So much for pseudo-problems and some of their gratuitous solutions. Before addressing 
some real problems I would like to express my deep and abiding admiration for the 
work of Professor d'Espagnat, whose numerous books and articles 
(e.g.,~\cite{dECF,dENTDR,dERP,dEVR,dESEP,dEIHR,dERILT}) demonstrate 
conclusively that no interpretation of QM that takes the quantum state for more than a 
probability measure is consistent with Option~A. It is not the intention of this 
Comment to belittle that outstanding achievement.

According to Feynman, the mother of all quantum effects is the two-slit experiment 
with electrons~\cite{Feynmanetal65}. If nothing indicates the slit taken by an electron 
then the electron goes through both slits without going through a particular slit and 
without having parts that go through different slits. How can this be? That's what I 
call a problem. To my way of thinking, the origin of the problem is a mismatch between 
the spatial aspect of the world and the way we all tend to think about it. Ask yourself 
how you think about the empty regions $L$ and $R$ inside the two slits. No doubt you 
will consider them different, separate, distinct. Yet if they were distinct then either the 
electron would go through a particular slit or it would be divided into parts by its 
passage through both slits. Hence if nothing indicates the slit taken by the electron then 
those regions can't be distinct for the electron, and so they can't be distinct {\it per se}. 
That's what baffles us.

But think again. How are $L$ and $R$ different? You may say, well, they are in 
different places; they have different positions. So where are these different places? 
Your answers will have the following form: ``$L$~is at/inside~$\cal L$'' and ``$R$~is 
at/inside~$\cal R$'', where $\cal L$ and $\cal R$ are shorthand notations for whatever 
you say the positions of $L$ and $R$ are. So where is $\cal L$---the position of $L$ or 
the region containing~$L$? It is clear that you are poised for an infinite regress. Your 
answer will have the form: ``$\cal L$~is at/inside~${\cal L}_1$'', where ${\cal L}_1$ is 
short for whatever position you attribute to~$\cal L$, and so on. 

The root of the problem is that we keep switching between two inconsistent modes 
of thinking. If we say ``$L$~is at/inside~$\cal L$'', we treat $\cal L$ as a property 
(namely, the position of~$L$) and we treat $L$ as a thing that has a property 
(namely, the position~$\cal L$). If we then ask ``Where is~$\cal L$?'', we treat $\cal L$ 
as a thing to which a position can be attributed. But we can't have it both ways. Either 
$\cal L$ is a thing to which properties (such as a position) can be attributed, or $\cal L$ 
is a position---a property---that can be attributed to things.

Hence our problem actually has an easy solution: Stop thinking of positions and 
regions of space as if they were {\it things}. $L$ and $R$ are {\it properties} that 
may or may not be possessed. $L$---that is, the property of being inside $L$---is 
possessed just in case there is a thing $T$ inside $L$---a thing that has the 
property~$L$. We tend to think that saying (i)~``$T$~is inside~$L$'' is different 
from saying (ii)~``$T$~has the property~$L$'' because proposition~(i) seems to 
imply that $L$ is a thing that contains~$T$. But this is where we are wrong. It is 
logically inconsistent to think of properties as if they were things. $L$ is not a 
thing, and proposition~(i) says exactly what proposition~(ii) is saying.

Once we stop thinking of positions and regions as if they were things, we are no 
longer bothered by the behavior of electrons in two-slit experiments. If an electron 
goes (as a whole) through both slits, neither $L$ nor $R$ is attributable to it, nor 
does it have parts to which $L$ and $R$ are separately attributable. What is 
attributable to the electron is $L\cup R$, and since this is not a thing that has $L$ 
and $R$ for its parts but a property, there is no reason in the world why the 
possession of the property $L\cup R$ should entail the possession of $L$, the 
possession of $R$, or the existence of parts possessing the respective properties 
$L$ and $R$---no reason other than the fallacy of thinking of space as if it were a 
thing that has parts. The behavior of electrons in two-slit experiments forces us to 
acknowledge a fallacy we have previously committed with impunity because the 
world of classical physics was consistent with it.

Once we commit this fallacy we end up thinking that any finite region of space has 
infinitely many distinct parts, and if we insist on thinking of {\it all} parts of space 
as self-existent and intrinsically distinct, we end up with the substantive, set-theoretic 
conception of space as a manifold of intrinsically distinct point 
individuals (usually considered in one-to-one correspondence with triplets of real 
numbers and denoted by \RRR). I am not advocating that we should stop using 
\RRR\ as a mathematical tool. But we must recognize it for what it is. We must 
learn to think of the elements of \RRR\ not as things that have positions but as 
positions that things may have. A ``coordinate point'' $\cal P$ and ``its position'' 
are the same animal. $\cal P$~{\it is} a position and therefore it does not {\it have} 
a position. Only material objects have positions, and only those positions that are 
actually possessed exist, and only those positions that are indicated (by facts) are 
actually possessed. The others exist solely in our imagination. It is therefore 
necessary to make a clear distinction between the set \RRR\ of all (exact) positions 
that a material object {\it may} have, and the spatial aspect of the world---the 
positions that are {\it actually} possessed by material objects. It won't do to 
regard \RRR\ itself as adequately representing the spatial aspect of the world.

Since no position is possessed unless it is indicated, and since nothing ever 
indicates an exact position, nothing ever has an exact position. A position 
measurement can never distinguish between more than a finite number of finite 
regions, and this is why attributable positions are always finite regions like 
$L$ and $R$. If we further take into account that in order to specify the position 
of an object we must say where it is in {\it relation} to another object that serves as 
a reference point, we arrive at the conclusion that space---the spatial aspect of the 
world---is the totality of {\it relative} positions (or spatial {\it relations}) that exist 
between material objects. This has a number of surprising 
consequences---surprising because they are at odds with our deep-seated 
misconceptions about the nature of space. For one, there is no such thing as 
``empty space''. If there are no objects, there are no spatial relations, and hence 
there is no space. A world without objects is a spaceless world. For another, there 
is no such thing as ``the form of an electron''.

What is clear right away is that if an object without parts---a fundamental particle 
like the electron---had a form then this could only be the form of a point. If it had 
any other form, it would have parts. Now try to imagine a single pointlike object. 
As you imagine a pointlike object, you also imagine a spatial expanse in which this 
object is situated. You cannot imagine a point without imagining a space that 
surrounds or contains it. A point is a form, and the existence of a form implies the 
existence of space. But you are asked to imagine a {\it single} pointlike object---not 
any other thing, nor any of this object's relations to other things. Therefore you 
must not imagine its external spatial relations. (The external spatial relations of an 
object $O$ are those between $O$ and objects that have no parts in common with 
$O$.) And since a pointlike object lacks parts and therefore lacks internal spatial 
relations, your mental picture must not contain {\it any} spatial relations. And 
since space consists of spatial relations, your mental picture must not contain 
space. And since the existence of a form implies the existence of space, your 
mental picture must not contain any form. The upshot is that the existence of a 
pointlike form is inconsistent with the proper way of thinking about space---as a 
set of spatial relations. A fundamental particle like the electron therefore is a 
formless entity.

A possible way of giving QM in a nutshell is to say that there are 
limits to the objective reality of our conceptual distinctions. If an electron as a 
whole goes through both slits then the distinction we make between ``The electron 
goes through~$L$'' and ``The electron goes through~$R$'' is a distinction that 
Nature does not make; it corresponds to nothing in the world; it exists solely in 
our heads. Here we are talking about spatial distinctions, but the same is true of 
our substantial distinctions. What the two-slit experiment is to spatial distinctions, 
a two-particle collision is to substantial distinctions. Suppose that initially we have 
two incoming particles, one heading northward and one heading southward, and 
that after the collision we have two outgoing particles, one heading 
eastward and one heading westward. QM tells us in unmistakable terms that if the 
particles are of the same type (and their spins are not antiparallel) then the 
outgoing particle heading eastward is neither the same as nor different from either 
of the incoming particles. The distinction between ``$E$ is identical with $N$'' 
and ``$E$ is identical with $S$'' (where $E$ stands for the outgoing particle 
heading eastward and $N$ and $S$ stand for the incoming particles) is another 
distinction that Nature does not make; it corresponds to nothing in the world; it 
exists solely in our heads.

If we try to think of the properties that a fundamental particle possesses ``by 
itself'', out of relation to other things, we find that there aren't any. The properties 
of a fundamental particle are either relational (like positions or momenta) or 
dynamical (characteristic of their interactions, like charges) or comparative (like 
mass ratios---the mass of a single particle has no physical significance). Can we 
nevertheless say that two fundamental particles, considered in themselves (and 
therefore out of relation to each other) are distinct? Can we say that they are {\it two}? 
According to the Identity of Indiscernibles, a principle of analytic ontology which 
says that two things cannot have exactly the same properties, the ``two particles'' 
considered in themselves are two things only if they possess the property which 
philosophers call ``thisness'' or ``haecceity''. But the possession of this property 
implies that the two particles in our collision experiment are re-identifiable, 
and QM makes it abundantly clear that they are not re-identifiable. Hence if two 
fundamental particles are considered in themselves they cease to be two. They become 
identical not just in the weak sense of exact similarity but in the strong sense of {\it 
numerical identity}.

Then what is this one and the same thing $X$ that every fundamental particle 
intrinsically is? Since considered in itself, out of relation to other things, a 
fundamental particle has no properties, all we can say of an (existing) fundamental 
particle {\it in itself\/} is that {\it it exists}. Hence that which every fundamental 
particle intrinsically is, is {\it existence pure and simple}. Let us call it ``Existence'' 
with an upper-case~E.

How is Existence related to space? Space contains---in the proper, set-theoretic 
sense of ``containment''---the forms of all things that have forms. It does not 
contain material objects over and above their forms; {\it a fortiori} it does not 
contain the formless ``constituents'' of matter. Space exists {\it between} the 
fundamental particles; it is spanned by their spatial relations. And since what 
exists at either end of each spatial relation is Existence, spatial relations are {\it 
internal} to Existence. QM tells us that the physical world is both constituted by 
Existence and suspended within it.

Ontologies tend to be modeled after the grammatical relation between a subject 
and a predicate, and matter tends to be identified with the ultimate 
subject---that which is the same in things with different properties, the 
grammatical subject by itself, bereft of predicates. But matter also tends to be 
thought of as that which is different in things with the same properties. We ordinarily 
proceed on the implicit assumption that identical things come equipped with ``thisness'', 
and for this property only matter itself can be responsible. This way of thinking is at the 
roots of the Platonic-Aristotelian dualism of Matter and Form and its subsequent 
transformations, including the preposterous field-theoretic notion that physical 
properties are instantiated by the ``points of space''~\cite{RedheadQFT}. 

The nonexistence of ``thisness'' forces us to look upon Existence, rather than upon 
the fundamental particles, as the ontological equivalent of the grammatical 
subject. The One is logically and ontologically prior to the Many, which come into 
being when formless Existence enters into spatial relations with itself and 
acquires, as a consequence, the aspect of a multiplicity of formless particles. Along 
with the particles, space and forms come into being, for space is the totality of 
existing spatial relations (between Existence and Existence) and forms are 
particular sets of such relations.

The relations are logically 
and ontologically prior to the relata---the fundamental particles. We are prone to 
hold the opposite view---that spatial relations are supported by a 
self-existent multiplicity. In reality the multiplicity is supported by relations, 
which are supported by Existence, which is one. QM describes a world that is 
created top-down, by a process of differentiation, rather than bottom-up, by a 
process of aggregation. (Saying that QM describes the world is very different from 
saying that some or all of the mathematical symbols of QM describe the world. It 
takes a considerable amount of thought to get from probability measures to their 
ontological import.)

The world is differentiated both spacewise (spatial relations warrant distinctions 
between ``here'' and ``there'') and timewise (temporal relations warrant 
distinctions between ``now'' and ``then''). The temporal differentiation is effected 
by {\it change}, for time and change are coimplicates: A timeless world cannot 
change, and a changeless world is temporally undifferentiated and therefore 
timeless. To my way of thinking, the quintessential message of QM is that there 
are limits to the world's spatial and temporal differentiation. The world is only 
finitely differentiated. In an infinitely differentiated world, spatial relations are 
determinate quantities; they possess definite values. In a finitely differentiated 
world, spatial relations are indeterminate quantities; they possess fuzzy values, 
and so do temporal relations~\cite{diff}. The proper conceptualization of 
indefiniteness requires the use of statistical concepts, and this is why QM is 
formally a statistical theory.

The proper way of dealing with fuzzy quantities is to make counterfactual 
probability assignments~\cite{Mohrhoff00}. If a quantity is said to have an 
``indefinite value'' what is really intended is that it {\it would} possess a value if it 
{\it were} successfully measured, and that at least two possible values have positive 
probabilities of being found. (The counterfactuality cannot be eliminated but it 
may be shifted from measurements to fuzzy values: If a measurement of 
observable $Q$ {\it is} successfully performed on an ensemble of identically 
prepared systems and the results have positive dispersion, the value of $Q$ {\it 
would} be fuzzy for an individual system $S$ if the measurement {\it were} not 
performed on $S$.)

So how do we get from fuzzy values or counterfactual statements to unconditional 
statements of value-indicating facts? No property is a possessed property unless it 
is an indicated property. This seems to entail a vicious regress, which at first blush 
looks like just another version of von Neumann's ``catastrophe of infinite regression''. 
The positions of detectors are extrinsic, too. They are what they are 
only because of the facts that indicate what they are. This requires the existence of 
detector detectors indicating the positions of particle detectors, which requires the 
existence of detectors indicating the positions of detector detectors, and so on {\it 
ad infinitum}. Generally speaking, the (contingent) properties of things ``dangle'' 
ontologically from what happens or is the case in the rest of the world. Yet what 
happens or is the case there can only be described by describing material objects, 
and their properties too ``dangle'' from the goings-on in the rest of the world. This 
seems to send us chasing the ultimate property-indicating facts in never-ending 
circles. Somewhere the buck must stop if the ontological story unfolding in this 
section is to be a viable interpretation of QM.

To begin with, the following points should be kept in mind. First, although the 
teapot isn't only there when somebody looks, it is there only because of the myriad 
of facts that betoken its presence. If there weren't any actual event or state of 
affairs from which its position could be inferred, it wouldn't have a position, or 
else its position wouldn't have a value. (There is no need for a conscious observer 
to actually carry out the inference.)

Second, as it stands the problem is still ill posed, for we do not proceed from 
counterfactuals to unconditional statements, nor do we start with valueless 
positions in search of value-indicating facts. We proceed from facts and the 
statistical correlations obtaining among them. These correlations warrant 
(i)~inferences to the existence of objects with kinematical properties that have 
fuzzy values and (ii)~the interpretation of the statistically correlated facts as 
indicating possessed values.

Third, the ``measurements'' to which both the minimal instrumentalist 
interpretation of QM and Jauch's definition of ``state'' refer, are not confined to 
manipulations that are {\it intended} to determine the value of a given observable 
or that lead to the acquisition of {\it knowledge}. The sufficient condition for a 
measurement is an actual event or state of affairs that warrants the assertability of 
a statement of the form ``$S$~has property $p$ at time~$t$'', irrespective of 
whether anyone is around to assert, or take cognizance of, that event or state of 
affairs, and irrespective of whether it has been anyone's intention to learn 
something about~$S$. Bohr insisted that quantum systems should not be thought 
of as possessing properties independently of experimental 
arrangements~\cite{dEBohr}. His insistence on the necessity of describing 
quantum phenomena in terms of experimental 
arrangements~\cite{BohrAPHK,BohrATDN} does not mean that quantum 
phenomena require the existence of experimental physicists. For ``experimental 
arrangement'' read: the totality of property-indicating facts. {\it Any} matter of 
fact that ``is about'' (has a bearing on) the properties of a physical system, 
qualifies as a measurement result.

Fourth, the extrinsic nature of the contingent properties of physical systems 
follows from the fuzziness of their values, inasmuch as this requires the use of 
counterfactual probability assignments~\cite{MohrhoffRK}. The use of 
conditionals with false antecedents would be gratuitous if the antecedents were 
never true, for in this case the conditionals could not be tested. But, in fact, the 
conditionals are abundantly tested, for they express the statistical correlations 
among facts that QM is concerned with, and no experiment or observation has 
ever given the lie to QM. This warrants the counterfactual use of the correlations 
(that is, it warrants the assignment of probabilities to the possible results of 
unperformed measurements), and this is the formal expression of 
indefiniteness. But if the antecedents of conditional probability assignments can 
be false as well as true, there has to be a criterion for when they are true, and this 
consists in the existence of value-indicating facts.

Value-indicating facts are actual events or states of affairs. Events are changes in 
the properties of objects; states of affairs concern the properties of objects. These 
properties are extrinsic; their possession is not factual {\it per se}. Yet facts are 
per definition factual {\it per se}. The task of resolving this apparent paradox is 
the genuine core of the ``measurement problem''.

The positional indefiniteness of an object $O$ finds expression in the 
unpredictability of the results of position measurements performed on~$O$. 
Evidence of the indefiniteness of $O$'s position, or of the corresponding statistical 
dispersion, requires the existence of detectors with sensitive regions that are small 
and localized enough to probe the range of values over which $O$'s position is 
distributed. (A detector is any object capable of indicating the presence of another 
object in a particular region of space.) The indefiniteness of $O$'s position cannot 
evince itself through statistically distributed 
position-indicating events if there are no detectors with sharper positions and with 
sensitive regions that are smaller than the space over which $O$'s position is 
distributed. But detectors with sharper positions and sufficiently small sensitive 
regions cannot exist for all detectable objects. There is a finite limit to the 
sharpness of the positions of material objects, and there is a finite limit to the 
spatial resolution of actually existing detectors. Hence there are objects whose 
positions are the sharpest in existence. These never evince their indefiniteness 
through unpredictable position-indicating events. Such objects are entitled to be 
called ``macroscopic''. We cannot be certain that a given object qualifies as 
macroscopic, inasmuch as not all matters of fact about its whereabouts are 
accessible to us, but we can be certain that macroscopic objects exist.

If the positional indefiniteness of a macroscopic object never evinces itself through 
unpredictable position-indicating events---the occasional unpredictability of the 
position of a macroscopic pointer reveals the indefiniteness of a property of 
another object, not the indefiniteness of the position of the pointer---then it is legitimate 
to ignore the positional indefiniteness of macroscopic objects. And if it 
is legitimate to ignore this---not only for all practical purposes but strictly---then it 
is legitimate to treat the positions of macroscopic objects as intrinsic.

The step from acknowledging the extrinsic nature of all contingent properties to 
treating the positions of macroscopic objects as intrinsic is of the same nature as 
the step from acknowledging the purely correlative character of classical laws of 
motion to the use of causal language. According to Hume~\cite{Hume}, causality is 
in the eye of the beholder; it is our way of interpreting events, not a feature of the 
events in themselves. QM has proved him absolutely right. Macroscopic objects 
evolve predictably in the sense that every time the position of such an object is 
indicated, its value is consistent with all predictions made on the basis of (i)~all past 
indicated properties and (ii)~the classical laws of motion. (As mentioned above, 
there is one exception: Whenever the position of such an object serves to indicate 
an unpredictable property of the quantum domain, it is itself not predictable.) 
This makes it possible to think of the positions of macroscopic objects as forming 
a self-contained system of positions that ``dangle'' causally from each other, and 
this makes it possible to disregard that in reality they ``dangle'' ontologically from 
(supervene on) position-indicating facts. The possibility of using causal concepts 
in the classical domain implies the possibility of treating the properties of the 
classical domain as intrinsic.

While correlations that are not manifestly indeterministic (like those between the 
successive positions of a macroscopic object) can be embellished with causal 
stories, in the quantum domain causal concepts are entirely out of 
place~\cite{Mohrhoff00}. Causality is a function of psychology, not of physics. It 
is rooted in our self-perception as agents in a successively experienced world. We 
can impose it on the classical domain with some measure of consistency, although 
this entails the use of a wrong criterion: Temporal precedence takes the place of 
causal independence as the criterion which distinguishes the cause from the effect. 
But when we deal with correlations that are manifestly indeterministic, projecting 
our agent causality into the physical world no longer works. Trying to causally 
explain these correlations is putting the cart in front of the horse. It is the 
statistical correlations that explain why causal explanations work to the extent they 
do. They work in the classical domain where statistical variations are not in 
evidence. In this domain we are free to use language suggestive of nomological 
necessity. But if we go beyond this domain, we realize that all correlations are 
essentially statistical, even where statistical variations are not in evidence, and that 
our belief in nomological necessity is just that---a belief.

By the same token, the possibility of treating the positions of macroscopic objects 
as intrinsic does not mean that they {\it are} intrinsic. The world is spatially 
differentiated to the extent that the values of spatial relations are indicated by 
facts, and facts never indicate numerically precise values. Even the positions of 
macroscopic objects are fuzzy~\cite{MF}; therefore even they are extrinsic. Yet 
their fuzziness exists only in relation to a backdrop that is more differentiated 
spacewise than is the actual world, a backdrop that exists only in our 
imagination. Space---so we must keep reminding ourselves---isn't an intrinsically 
and infinite differentiated container of objects. It is a set of more or less fuzzy 
relations. Some of these relations---those that obtain between macroscopic 
objects---are the sharpest in existence, and these are not fuzzy in any real sense; they are 
fuzzy only in relation to an unrealized degree of spatial differentiation. Facts that 
are indicative of the positions of macroscopic objects are correlated in such a way 
(viz., predictably) that any reference to them is superfluous: We can think of these 
positions as intrinsic and as evolving deterministically, and this for all {\it 
quantitative} purposes rather than merely for all ``practical'' ones. The positions of 
macroscopic detectors are not truly sharp, but as they are the sharpest in existence we 
may treat them as sharp, as intrinsic, and hence as {\it per se} available as possible 
properties of (things in) the quantum domain.

Recall that attributing factuality (whether to a nomologically possible world or to a 
measurement outcome) is beyond the scope of any theory. When the theory has 
done its part, we are left with the problem of attributing factuality. And this 
problem has exactly one consistent solution. The unaccountable factuality of facts 
belongs to those properties which can be treated as intrinsic because their 
indefiniteness exists solely in our heads.

\section{\large REALITY VEILED AND\\
UNVEILED}

Bohr felt that our interpretational difficulties ``hardly allow us to hope that we 
shall be able, within the world of the atom, to carry through a description in space 
and time that corresponds to our ordinary sensory perceptions''~\cite{Bohr23}. 
While this is past the shadow of a doubt, Stapp's conclusion that ``$\,$`space,' like 
color, lies in the mind of the beholder''~\cite{Stapp} is a {\it non sequitur}. This 
would follow if a spatiotemporal description corresponding to our sensory 
perceptions were the only possible description. But, as the previous section 
has shown, an entirely different account of the spatiotemporal aspect of the world 
is (i)~possible and (ii)~consistent with both standard QM and Option~A.

There is mounting evidence from neuroscience that visual perception and visual 
imagination share the same processing 
mechanisms~\cite{Finke,FinkeShepard,ShepardCooper}. Hence as long as we 
insist that science (which is in the business of constructing theories or {\it 
conceptual} models of the world) provide us with a model that can be {\it 
visualized}, we limit its scope by the very brain mechanisms that are instrumental 
in the construction of the phenomenal world---the world as we humans perceive it. 
This is precisely what Option~B does. D'Espagnat seems to think that such a limit 
to the scope of scientific inquiry is unavoidable, and he tries to make a virtue of 
this perceived necessity.
\begin{quote}
I have here in mind a viewpoint that would be totally faithful to the\ldots scientific 
ideal of keeping to what seems unquestionable within collective human experience, 
namely the impressions we share, without any admixture of presupposed ideas 
concerning the actual existence of the forms thus perceived\ldots
\end{quote}
The way in which the brain processes visual information guarantees that the 
result---the phenomenal world or ``the impressions we share''---is a world of 
objects that are bounded by surfaces~\cite{CCP,BCCP}. The phenomenal world 
conforms to the ``cookie cutter paradigm'' (CCP) according to which the world's 
synchronic multiplicity rests on surfaces that carve up space in the manner of 
three-dimensional cookie cutters: The parts of any material object are defined by 
the parts of the space it ``occupies'', and the parts of space are defined by 
delimiting and separating surfaces. This seems self-evident because this is how we 
perceive the world because this is how the brain analyses visual information.

As long as we take the parts of matter to be defined by the parts of space, the 
parts of space are logically prior to the parts of matter; hence they exist 
independently of matter; hence space is a thing that has parts. Another, probably 
similarly hard-wired misconception is that parts exist by themselves (rather than 
by virtue of some process of division or differentiation) or that multiplicity (rather 
than One Existence) is fundamental. Combining these misconceptions leads to 
the idea that all (conceivable) parts of space exist by themselves and to the concept of 
space as a manifold of intrinsically distinct point individuals.

If the world were created along the lines laid down by the CCP, the shapes of 
things would be bounding surfaces, and matter would be an extended stuff 
bounded by surfaces. A material object would have as many parts as the space it 
occupies, and an object without parts---a particle like the electron---would be a bit 
of stuff with the form of a point. Extended material objects would always occupy 
distinct parts of space, and the positions of pointlike objects would always be 
distinct. Material objects would be re-identifiable since at every time there would 
be a fact of the matter concerning which is which.

If we subscribe to the ``scientific ideal of keeping to\ldots the impressions we 
share'', we remain committed to the CCP, and this, clearly, leads us up the garden 
path. It implies the substantive conception of space as a manifold of intrinsically 
distinct point individuals while QM tells us that space is a set of relations between 
material objects. It implies that the shapes of things are points or bounding 
surfaces while QM tells us they are sets of spatial relations. It implies that 
electrons are pointlike while QM tells us they are formless. It implies that particles 
are re-identifiable while QM tells us they are not. It further implies that the world 
is infinitely differentiated spacewise and timewise while QM tells us that it is only 
finitely differentiated; that spatial relations are determinate quantities while QM 
tells us they are fuzzy; that the world is created bottom-up by aggregation while QM 
tells us it is created top-down by differentiation.

The upshot is that {\it QM is inconsistent with Option~B}. Option~B commits us 
to the CCP, and every one of the implications of the CCP directly contradicts 
what QM is trying to tell us. D'Espagnat concludes that we must either accept 
breaks in the linear evolution of quantum states or
\begin{quote}
grant that man-independent reality\ldots is something more ``remote from 
anything ordinary human experience has access to'' than most scientists were up to 
now prepared to believe\ldots
\end{quote}
I fully agree with this conclusion. The world according to QM, as outlined in 
Sec.~2, is more remote from anything ordinary human experience has access to 
than most scientists were up to now prepared to believe. But this does not mean 
that we cannot understand it. I entirely disagree with his claim that
\begin{quote}
while, through physics, Being informs us quite definitely of what it is not\ldots it 
seems reluctant at letting us know what it truly is.
\end{quote}
Nothing could be further from the truth. Reality does not veil itself~\cite{dEVR}. 
It is {\it we} who veil it, by clinging to (i)~concepts of space, time, form, and 
substance that are inapplicable to the physical world and (ii)~pseudo-realistic 
ways of thinking about probability measures. Once these misconceptions are 
replaced by adequate ways of thinking, everything is above board. Nothing 
remains mysterious, except the mother of all mysteries---why there is anything at all, 
rather than nothing. As Wittgenstein said in the {\it Tractatus Logico-Philosophicus\/}: 
``Not how the world is, is the mystical, but {\it that} it is." QM refers to this mystery 
twice: when it presupposes the unaccountable 
factuality of facts, and when it tells us that intrinsically each fundamental particle 
is existence pure and simple. These {\it aper\c cus} of ``bare reality'' 
play distinct ontological r\^oles. While the factuality of facts is the ultimate reason 
why there are properties, Existence is the ultimate reason why there are things that 
have properties.


\begin{thebibliography}{99}
\bibitem{Time}
Times are indicated by the positions of (macroscopic) hands of clocks. 
Macroscopic objects are defined in Sec.~2.
\bibitem{diff}
In a world that is infinitely differentiated spacewise, all objects have exact 
positions. In such a world a moving object passes through infinitely many distinct 
(actual) states in any finite time span. Hence such a world is infinitely 
differentiated timewise. In a world that is only finitely differentiated spacewise, no 
object passes through infinitely many distinct actual states in a finite time span. 
Hence such a world is only finitely differentiated timewise.
\end{thebibliography}
\end{document}